\documentclass{cernyrep} 
\usepackage{texnames}
\usepackage[T1]{fontenc}
\begin{document}
\title{Heavy Ion Collisions at the dawn of the LHC era}

\author{J.~Takahashi}

\institute{Universidade Estadual de Campinas, S\~{a}o Paulo, Brazil}

\maketitle 
 
\begin{abstract}
This is a proceeding of the CERN Latin American School of High-Energy physics that took place in the beautiful city of Natal, northern Brazil, in March 2011. In this paper I present a review of the main topics associated with the study of Heavy Ion Collisions, intended for students starting or interested in the field. It is impossible to summarize in a few pages the large amount of information that is available today, after a decade of operations of the RHIC accelerator and the beginning of the LHC operations. Thus, I had to choose some of the results and theories in order to present the main ideas and goals. All results presented here are from publicly available references, but some of the discussions and opinions are my personal view, where I have made that clear in the text.
\end{abstract}

\section{Introduction}

In this very exciting field of science, we use particle accelerators to collide heavy ions such as Lead or Gold nuclei, instead of colliding single protons or electrons. By doing so, we produce a much more violent collision where a large number of particles are created and a considerable amount of energy is deposited in a volume bigger than the size of a single proton. As a result, a highly excited state of matter is created and this state can have different characteristics from regular hadronic matter. It is postulated that if the energy density is high enough, the formed system will be in a state where quarks and gluons are no longer confined into hadrons and thus exhibits partonic degrees of freedom~\cite{qcd1,qcd2}. Since the main theory that explains the interaction of matter in these extreme conditions is the theory of Quantum CromoDynamics (QCD), by studying the system formed in these relativistic heavy ion collisions, we can explore the QCD phase diagram and understand the characteristics of the different phases of matter. 

In particular, QCD predicts that at extreme conditions, in high temperature or high baryoninc density, a new phase of matter known as the Quark-Gluon Plasma (QGP) would be formed, where partonic degrees of freedom could be observed in a volume larger that the size of a single hadron ~\cite{qgp1,qgp2}. On one extreme of the QCD phase diagram, where density is high and temperature is low, the reduction of the coupling constant at small distances would make the quarks and gluons behave as free partons, thus forming a deconfined state of partons. This kind of matter could exist at the center of very dense astrophysical objects such as neutron stars~\cite{QGPastro1}. On the other extreme, of very high temperature when the energy density exceeds some typical hadronic value, (~$1 GeV/fm^{3}$), matter would also go through a phase transition and form a deconfined state of quarks and gluons. Lattice QCD calculations~\cite{LatQCD1} predicts a phase transition to a quark-gluon plasma at a temperature of approximately T$\approx$ 170 MeV, which is equivalent to $\approx 10^{12}$ K. This extreme condition is believed to be similar to the early stages of the evolution of our universe just after the Big-Bang, thus, studying the characteristics of the QGP and how it evolves allows us probe the different stages of our universe expansion.

Of course, since we are colliding hundreds of nucleons at the same time and in each reaction thousands of different particles are produced, the observation and analysis of these events require some work. Moreover, most of the particles that we measure are from the final stages of the system evolution, after it has gone though the phase transition back into ordinary matter and suffered multiple scatterings, so these particles do not carry direct information from the partonic phase that we are interested in. But exactly because of this challenging task imposed by the complexity of a heavy ion collision and the subsequent dynamical evolution of the system formed, this field is, in my opinion, one of the most exciting and interesting fields of science that requires creative approaches for analysis, the study and integration of different topics and theories and an open mind for new and unexpected physics. Just as a quick example, after the first years of RHIC data taking (Relativistic Heavy Ion Collider at BNL)~\cite{RHICwhitepapers}, it was concluded that the system formed in the collisions of Gold nuclei at an energy of 200 GeV had characteristics very similar to a perfect liquid~\cite{perfectLiquid}, where hydrodynamic models and predictions could be used to explain some of the experimental observables. Before the start of the RHIC operations, most of us expected to find a gas like system where quarks and gluons would be free, but did not expect that the system would show such a strong collective behavior, and much less that these quarks and gluons would still interact strongly. In addition to hydrodynamical models~\cite{hydro}, statistical thermodynamic models~\cite{thermal} are also used to study the global characteristics of the system formed, while phenomenological models such as coalescence models are also used to study specific experimental observables. More recently, the first observation of heavy anti nuclei such as the anti-alphas~\cite{starAAlpha} and anti-hyper-tritons~\cite{starAHT} in heavy ion collisions at RHIC and also at the LHC (Large Hadron Collider at CERN) opens further possibilities for new physics topics to be studied in these collisions. Hence, the study of heavy ion collisions is a rich environment where many different models and theories can be tested.

In this paper, I will discuss some of the general signatures of the search for the QGP from the early days before the start of the RHIC accelerator, what was learned at RHIC and what is expected to be measured at the LHC.

\section{How to study these complex collisions ?}

We learn in the Rutherford scattering the importance of the impact parameter in a collision, which corresponds to the transverse distance between the centers of the colliding nuclei and determines the scattering angle. In heavy ion collisions, the impact parameter is also important because it will determine the size of the overlapping area between the two colliding nuclei, and hence, the initial size of the system formed. In very peripheral collisions, where the impact parameter is large, the overlapping region is very small. When the impact parameter is zero, the two nuclei will collide head on (also referred to as a central collision), and the overlapping area will be equivalent to the transverse area of the nuclei. 

Very much like in the Rutherford experiment, despite the great precision that can be achieved in the beam position, we cannot choose the impact parameter of the collision. However, we can try to determine an experimental observable that is directly connected to the impact parameter. Since the impact parameter is inversely related to the size of the overlapping area and thus to the amount of energy that will be transferred from the beam rapidity to the central rapidity region, it can be correlated to the number of particles produced in the central rapidity region. So, for example, by measuring the number of charged particles produced in the central rapidity region, also known as the multiplicity of an event, it is possible to determine the impact parameter of the reaction and thus determine the centrality of the collision. The exact relation between the measured multiplicity and the actual impact parameter can only be determined by comparing the experimental data to geometrical models that consider a nuclear profile and take into account the inelastic cross section of nucleus-nucleus collisions. Optical model approximation or Monte Carlo based Glauber models~\cite{glauber} are used to calculate the centrality of the collisions classified by the charged particle multiplicity. 

Another way to determine the impact parameter is to measure the number of nucleons that did not participate directly in the overlapping region, also known as spectator nucleons. They can be measured by detectors such as calorimeters placed in the same direction of the colliding beam (Zero Degree Calorimeters, ZDC), but after the dipole magnets that curves the trajectory of the non-interacting beam. In very central collisions (small impact parameter) there will be very little spectator nucleons whereas in very peripheral collisions (large impact parameter) there should be a high number of spectators. So, an anti-correlation between the signals measured in ZDC detectors with the charged particle multiplicity measured in the central rapidity region will be observed. A typical example of a charged particle distribution measured in Au+Au collisions at 200 GeV is presented in figure 1, extracted from Ref.~\cite{RDSPhD}. The blue line represents the experimental measurement of the frequency of events ($d\sigma/dN_{CH}$) that were measured as a function of the number of charged particles ($N_{CH}$). On the top part of the plot, the equivalent number of average number of participants ($\langle N_{part}\rangle$) and the corresponding impact parameter (b) calculated using a Monte Carlo Glauber model is presented.

\begin{figure}
\centering\includegraphics[width=.6\linewidth]{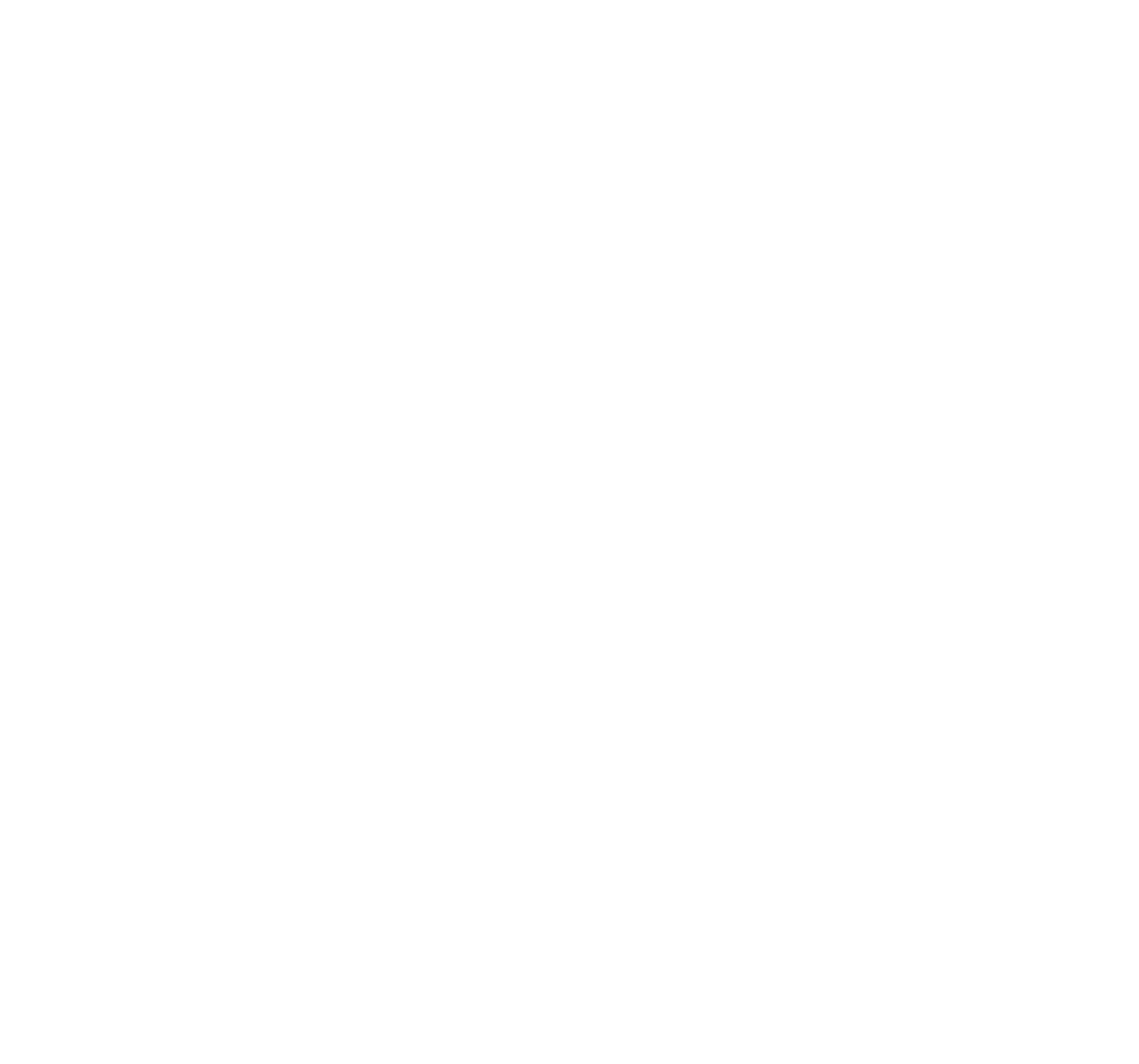}
\caption{ A typical example of a charged particle distribution measured in Au+Au collisions at 200 GeV, extracted from Ref.~\cite{RDSPhD}.}
\label{fig:1}
\end{figure}

In summary, it is possible to classify the measured events into different event centrality classes using the charged particle multiplicity, and to associate an average impact parameter to these events. By doing so, it is possible to study different observables as a function of the collision centrality, and thus, to study the dependence with the size of the system formed. Also, the possibility to separate data into different collision centrality classes is a very useful tool, since, a very peripheral collision should have similar characteristics as p+p collisions, where it is presumed that the formation of a QGP is not possible due to the small size and duration of the system. Hence, by comparing results from central collisions to peripheral or elementary collisions, it is possible to try to identify some unique signature of the Quark Gluon Plasma.

\section{The early signatures of the QGP search}

Early proposed signatures of the Quark Gluon Plasma were focused on the search for the phase transition by observing some sharp variation in the behavior of a physical parameter with increasing collision energy or increasing system size. Signatures related to the expected effects caused by the characteristics of the new state of matter to the final state particles were also proposed. Here I will discuss only briefly three of the signatures proposed, but there were many other observables proposed and searched for over the years. A more complete discussion on the proposed signatures of the QGP can be found at Ref.s~\cite{physicsQGP}.

It is expected that the deconfinement from the hadronic matter to the Quark-Gluon Plasma would be accompanied by chiral symmetry restoration. Hence, particles produced in the QGP would have their masses altered. If then we looked at short-lived resonance particles that would be formed and decayed within the plasma, the daughters of these particles would carry the information of the alteration caused by the chiral symmetry restoration. The natural place to look for these particles would be to observe the decay channels of vector mesons into weakly interacting electromagnetic probes such as leptons and photons since due to their low interaction cross section, they could leave the medium undisturbed and thus keep the original information of the parent particle. But the experimentally measured electron spectra have not only the contribution from the resonance decays, but also many other electrons from other sources, such as the decay of $\pi^{0}$ and $\eta$ mesons and also from gamma conversions. In figure 02, left panel, the plot obtained from Ref.~\cite{ceres2} shows the inclusive $e^{+}e^{-}$ invariant mass spectra measured in proton-Beryllium collisions by the CERES experiment at the SPS accelerator. In this plot, one can see the contributions from the different sources to the electron mass spectra and also that the sum of the different contributions adds to the total measured spectra, shown by the solid circles. This is a proof of principle measurement that shows that it is possible to identify all the different sources that contribute to the measured inclusive $e^{+}e^{-}$ invariant mass spectra. In the same figure, the $e^{+}e^{-}$ invariant mass spectra from the same experiment but measured in Lead-Gold collisions it is presented on the right panel, obtained from Ref.~\cite{ceres2}. The solid circles represent the experimental data points and the solid line represents the sum of all contributions from known sources of electrons. Clearly there is an excess of electron pairs in the low mass region, between 300 MeV/$c^{2}$ and 700 MeV/$c^{2}$. Also they find that this enhancement increase with collision centrality. It has been argued that this excess of electrons in the low momentum region could be due to thermal electrons from the Quark-Gluon Plasma itself, or a broadening of the vector meson mass peaks due to in medium modification caused by chiral symmetry transition at the phase boundary between the QGP and the hadronic matter~\cite{massShiftTheory}. Theoretical calculations suggest that the most appropriate for the observation of direct thermal electrons would be in a higher mass range than the observed excess. Some models considering in-medium modification of the vector mesons were able to reproduce the observed excess seen by the SPS experiments. This excess at the same mass region was also observed at higher energy Au+Au collisions at RHIC~\cite{phenixPRC81}, but the models used to describe the excess observed in the SPS data fail to describe the RHIC data. Direct measurement of a shift in the mass peak of the $\rho^{0}$ meson in the hadronic decay channel has also been reported at RHIC, and presented by the STAR experiment in Ref.~\cite{STAR_rho}. The observed modification of the meson mass has also been proposed as a possible signal of chiral symmetry restoration~\cite{Rapp}, but other effects such as re-scattering and Bose-Einstein correlations can also modify the final mass peak position and shape, so there are still discussions on the origin of the observed shift in the mass peak position. We hope that the new measurements from Pb+Pb collisions at LHC will bring new light into this puzzle. With the higher energy in the collision, it should be possible to extend the mass range of the di-lepton pair spectra and also relative contributions from direct QGP thermal radiation should be higher.

\begin{figure}
\centering\includegraphics[width=.9\linewidth]{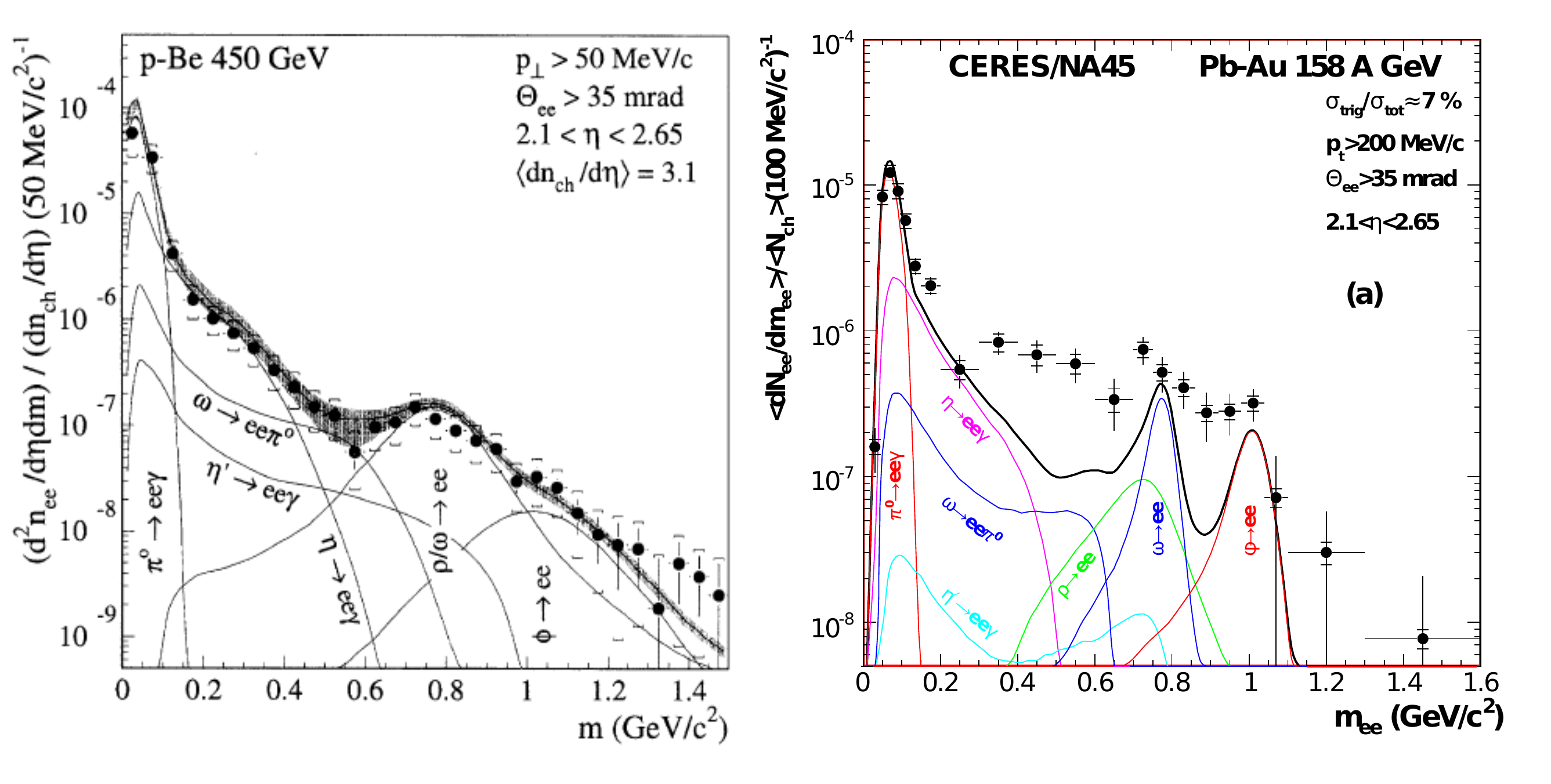}
\caption{Di-lepton invariant mass spectra, measured by the CERES experiment in p-Be collisions with beam energy of 450 GeV (left panel), and Pb-Au (right panel), with beam energy of 158 GeV. Figures were extracted from Ref.~\cite{ceres1} and~\cite{ceres2} respectively.}
\label{fig:2}
\end{figure}

Another important observable proposed as one of the main signatures of the QGP is the charmonium suppression~\cite{CharSup}. Due to the high density of quarks in the medium and a mechanism known as Debye screening, a pair of charm quark and charm anti-quark would have difficulty to bind and form a $c\bar{c}$ state, hence, resulting in a decrease of the J/$\Psi$ meson production in the QGP. But other processes will also affect the final measured yields of J/$\Psi$, such as the increase of hadronic absorptions and hadronic interactions (re-scattering and re-combination). Figure 03, left panel shows a compilation of J/$\Psi$ yields relative to Drell-Yan continuum measured in different collision systems, normalized by the expected rate of hadronic interactions as a function of system size (L represents the average path length that the $c\bar{c}$ pair traverse inside the medium after its creation). Figure extracted from Ref.~\cite{Schukraft}. The modeled hadronic interactions describe well the exponential decay observed in the J/$\Psi$ yields in smaller systems and peripheral Pb+Pb collisions, but fail to describe the yields at central collisions. A clear indication of an onset of suppression in the J/$\Psi$ yields can be observed starting for path length around 7 fm. Surprisingly, at collision energies measured at RHIC, similar levels of suppression were observed. A larger suppression was expected due to the higher energy density achieved at RHIC as expected by the original Debye screening mechanism. However, when going to higher collision energies, there is competition between suppression and recombination mechanisms and also an increase of feed-down from higher mass states that decay into J/$\Psi$. To identify the different contributions from these mechanisms one has to study the various dependencies of the suppression and compare to predictions from phenomenological models. In figure 03 right panel, the ratio of the J/$\Psi$ production between p+p and Cu+Cu measured by the STAR and PHENIX experiments is presented as a function of $p_{T}$. Figure extracted from Ref.~\cite{prc80_041902}. The suppression vanishes at high-$p_{T}$ ($p_{T} >$4 GeV/c), where the yield measured in central Cu+Cu collisions seems to be equivalent to a superposition of p+p collisions. Also shown in the same figure, different predictions for this ratio from phenomenological models. Details of the models and the discussions can be found at Ref.~\cite{prc80_041902}. In summary, the measurement that was originally expected to be a clean and clear signature of the QGP also needs to be studied carefully, where a detailed systematic study considering results from different collision energies and different system sizes is required. New measurements from the ALICE experiment at the LHC already show a suppression of the J/$\Psi$ produced in Pb+Pb collisions with respect to equivalent normalized p+p collisions~\cite{aliceJPsi}, but the final results and discussions are yet to be presented.

\begin{figure}
\centering\includegraphics[width=.9\linewidth]{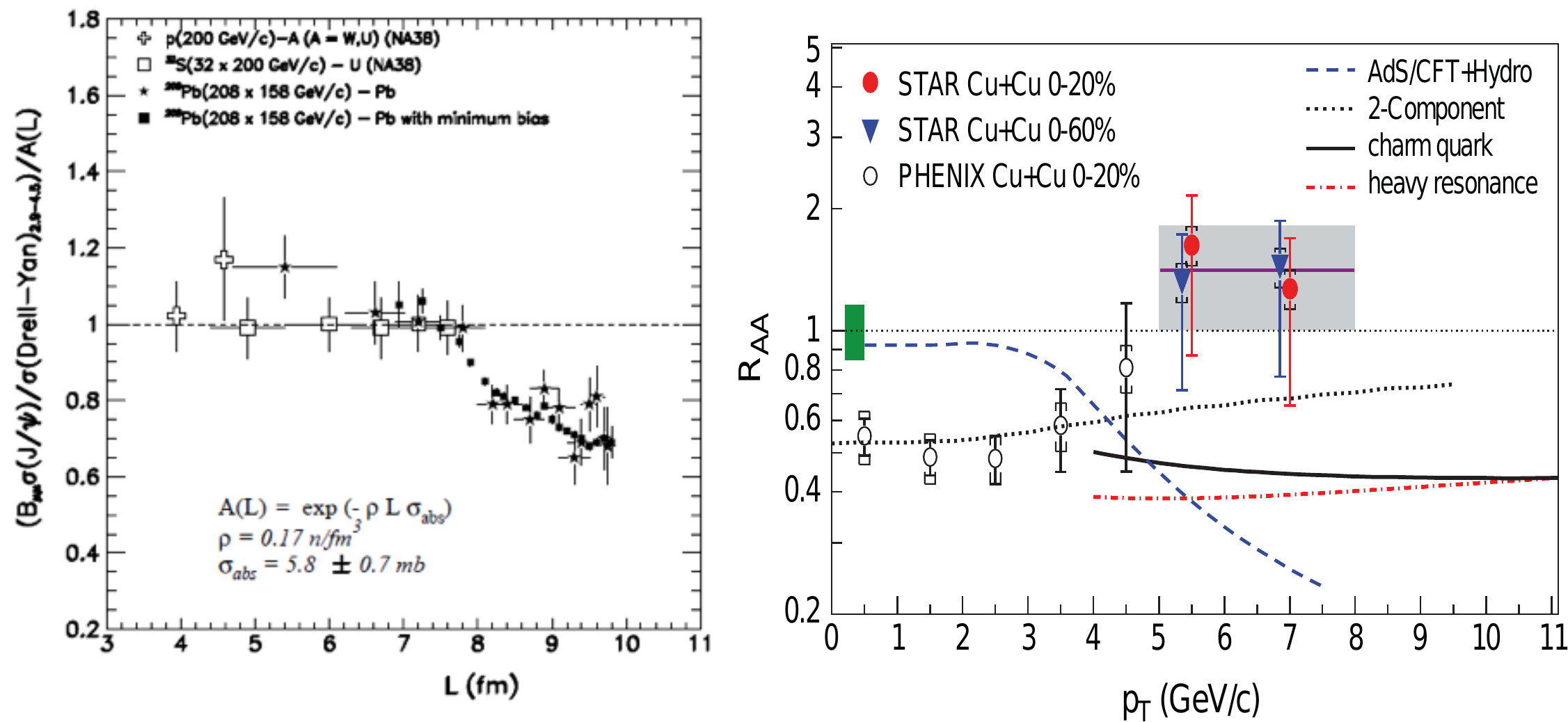}
\caption{Right panel: J/$\Psi$ yields relative to Drell-Yan continuum measured in different collision systems, normalized by the expected rate of hadronic interactions as a function of system size L (fm). Figure extracted from Ref.~\cite{Schukraft}. Left panel: Ratio of the J/$\Psi$ production between p+p and Cu+Cu measured by the STAR and PHENIX experiments as a function of $p_{T}$. Figure extracted from Ref.~\cite{prc80_041902}.}
\label{fig:3}
\end{figure}

To complete the discussion on the original proposed signatures of the QGP, we review the strangeness enhancement measurement. In a deconfined state, due to the decrease in the effective mass of the strange quarks and increase of $s\bar{s}$ pair production from partonic processes such as gluon-gluon fusion, an enhancement of strange particle productions is expected to be observed~\cite{strange}. In particular, if one assumes a rapid hadronization process after the QGP phase, the abundance of multi-strange particles such as $\Xi$ and $\Omega$ should be considerably greater in a reaction where QGP is formed than in a reaction with pure hadronic phase. Also, since the mass of the strange quark is of the order of the critical temperature, it is an excellent probe to test the thermalization of the system around the phase transition. By comparing the strange baryon yields measured in central Pb+Pb collisions to yields measured at elementary p+p and p+Be collisions, SPS experiments NA49 and NA57 observed an increase of the production higher than expected from simple scaling considering the volume of the formed system. Strange baryon yields were normalized by the number of participating nucleons (also known as number of wounded nucleons) to account for the normal increase due to the larger volume created in Pb+Pb collisions. In figure 04, extracted from Ref.~\cite{WWND2008}, we show the typical strangeness enhancement measurements that includes data from both RHIC-STAR experiment and SPS-NA57 experiment~\cite{Na57Strangeness}, for $\Omega$, $\Xi$ and $\Lambda$ baryon yields. The observed relative enhancement can also be interpreted as a effect of suppression of strange particle production in elementary collisions, due to canonical suppression. Within this picture, the smaller enhancement observed at higher collision energy can be well described considering that at RHIC energies, the strange particle productions in p+p collisions would be less suppressed. Also, the observed hierarchy of the enhancement between the different strange baryons can be understood within the context of canonical suppression of p+p collisions. In figure 04 we also show the $\phi$-meson enhancement presented by the STAR experiment in Ref.~\cite{STARphi}, from Au+Au collisions at 200 GeV. The $\phi$-meson enhancement contradicts the canonical suppression picture, for it is formed by a $s\bar{s}$ pair and thus should not be affected by the canonical suppression. So, in summary, the observed enhancement of strange baryons might be a result of both effects, a real enhancement of the strangeness production in heavy-ion collisions and suppression due to the limited size of the system in p+p collisions. Measurements from the LHC at higher energies will also be very important to elucidate this point, since lower canonical suppression is expected at higher energies, but also a higher enhancement might be present due to longer living system created in Pb+Pb collisions. Measurements of strange baryons in p+p collisions at energies of 900 GeV and 7 TeV has already been presented by several LHC experiments~\cite{LHCstrange}, and preliminary results presented only in conferences already show that strangeness enhancement is still observed at the LHC energies. More detailed results and the specific dependencies for the different particles species are yet to be presented.

\begin{figure}
\centering\includegraphics[width=.6\linewidth]{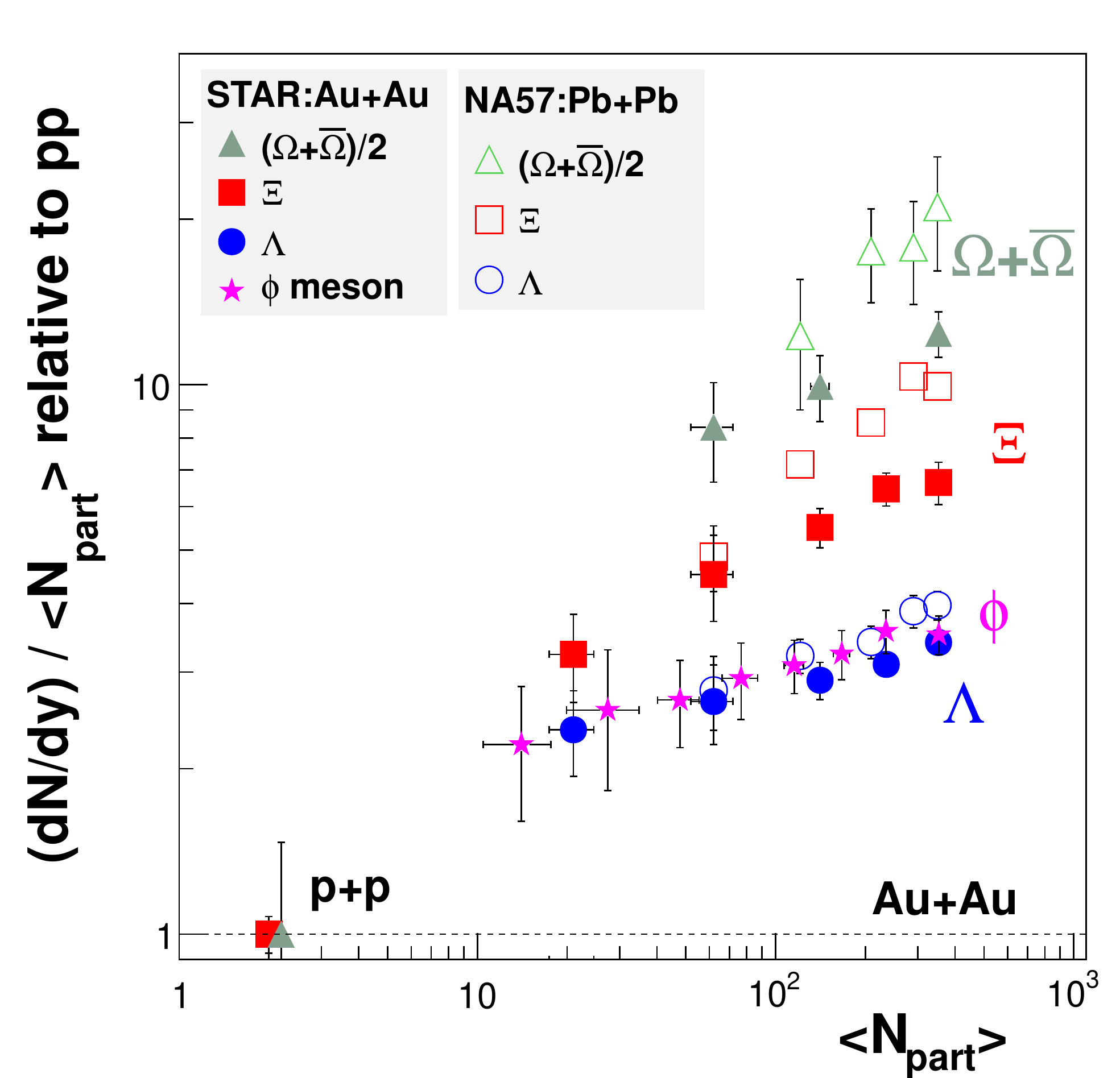}
\caption{Strangeness enhancement plot, with the relative yields of $\phi$, $\Lambda$, $\Xi$ and $\Omega$ of A+A collisions compared to elementary collisions, from RHIC and SPS data, as a function of the system size. Figure extracted from Ref.~\cite{WWND2008}. Plot shows the enhancement of the strange particle production with clear increase with system size.}
\label{fig:4}
\end{figure}

\section{New signatures observed at RHIC}

One of the most surprising and unexpected measurements observed at RHIC was the suppression of particle jets in central Au+Au collisions. In high energy collisions, particle jets are produced in hard processes when a large amount of momentum is transferred in the parton-parton interaction. The scattered partons will generate highly collimated and energetic particles known as jets. For this reason, the high momentum region of the transverse momentum spectra ($p_{T}>$4 GeV/c) is attributed to particles generated in hard scattering jets. Due to momentum conservation, these jets are usually produced in pairs and in opposite azimuthal directions. There are some QCD processes where three or more jets can be formed, but the back-to-back di-jet type events are the most common. 

Particle spectra measured in central Au+Au collisions are compared to the equivalent spectra from p+p collisions by a ratio called nuclear modification factor, also known as $R_{AA}$ ratio. The equivalent number of binary collisions normalizes each spectrum in the ratio to account for the fact that in heavy ion collisions there are more particles produced due to the larger number of interacting nucleons. In figure 05 left panel, we show the nuclear modification factor presented by the STAR experiment from Au+Au collisions at 200 GeV. Data from peripheral collisions (open blue circles) show that indeed, the high momentum region of the spectra is in agreement with the particle spectra measured in p+p collisions, where the ratio is consistent with one. However, the ratio from very central Au+Au collisions (solid red circles) shows that the spectrum at high transverse momentum is suppressed down to approximately $20\%$. This suppression of high momentum particles was attributed to absorption of the jets, in a very dense medium. Plot was generated using published data from Ref.~\cite{starRaa}. This interpretation was further corroborated when the same STAR experiment presented measurements of two particle azimuthal correlation showing the suppression of the back-to-back correlation in data from very central collisions, as shown in figure 05 right panel, extracted from Ref.~\cite{starBBCorrelation}. Within this interpretation, one of the hard scattered parton lost its energy by traversing a very dense medium and was absorbed, while the other parton that scattered to the opposite direction did not traverse the medium, hence fragmented into a jet of particles that was measured normally in the detectors. So, in the di-hadron azimuthal correlation function, only one jet is measured in central Au+Au collision (in $\Delta\phi\sim$0), while in p+p and peripheral collisions, a double jet structure (in $\Delta\phi\sim$0 and $\Delta\phi\sim\pi$) is observed. Different models and calculations show that indeed, if this was the case, the gluon density of the medium traversed by the parton would have to be very high, well above the predicted critical density for a phase transition. Since then, many other more detailed measurements further completed this scenario, such as the identified particle $R_{AA}$ plots, or variation of back-to-back jet suppression dependent on the azimuthal angle with respect to the reaction plane. The PHENIX experiment at RHIC also presented the equivalent $R_{AA}$ plots for direct photons~\cite{PhenixPhotons}. No suppression was observed, which is in agreement with the jet absorption picture since the photons would not interact strongly with the medium and thus should not be absorbed. More recent measurements of electron $R_{AA}$ from both STAR and PHENIX experiments at RHIC also showed a strong suppression at high-$p_{T}$. This comes as a surprise since, in the high-$p_{T}$ region, most of the electrons are supposed to be from decay of heavy-flavored particles. This would then indicate that even the very heavy charm and bottom quarks would be losing energy in the dense medium. A complete understanding of this measurement requires the separation of charm and bottom contributions to the inclusive electron spectra.

At the LHC, the ALICE experiment has also showed that the suppression in the high momentum region is still present even at collision energy of 2.76 TeV. In figure 06, extracted from Ref.~\cite{aliceRaa}, it is seen that the suppression measured in central Pb+Pb collisions is even larger than observed by the STAR experiment, suggesting that the medium formed at the LHC is denser than the one formed at RHIC. The higher $p_{T}$ range achieved by the ALICE data also allows us to observe the increase of the ratio, from $p_{T}$ 7 GeV/c to 20 GeV/c, which suggests that the suppression for more energetic partons is decreasing. Perhaps, with some more statistics, it will be possible to observe the limit where no more suppression would be observed. Also, more detailed measurements such as identified particle $R_{AA}$ and two particle correlation functions should help to further complete the picture and bring new ideas on how to probe the formed medium. CMS experiment also showed similar results consistent with jet quenching in Pb+Pb collisions~\cite{cmsJetQ}.

\begin{figure}
\centering\includegraphics[width=.9\linewidth]{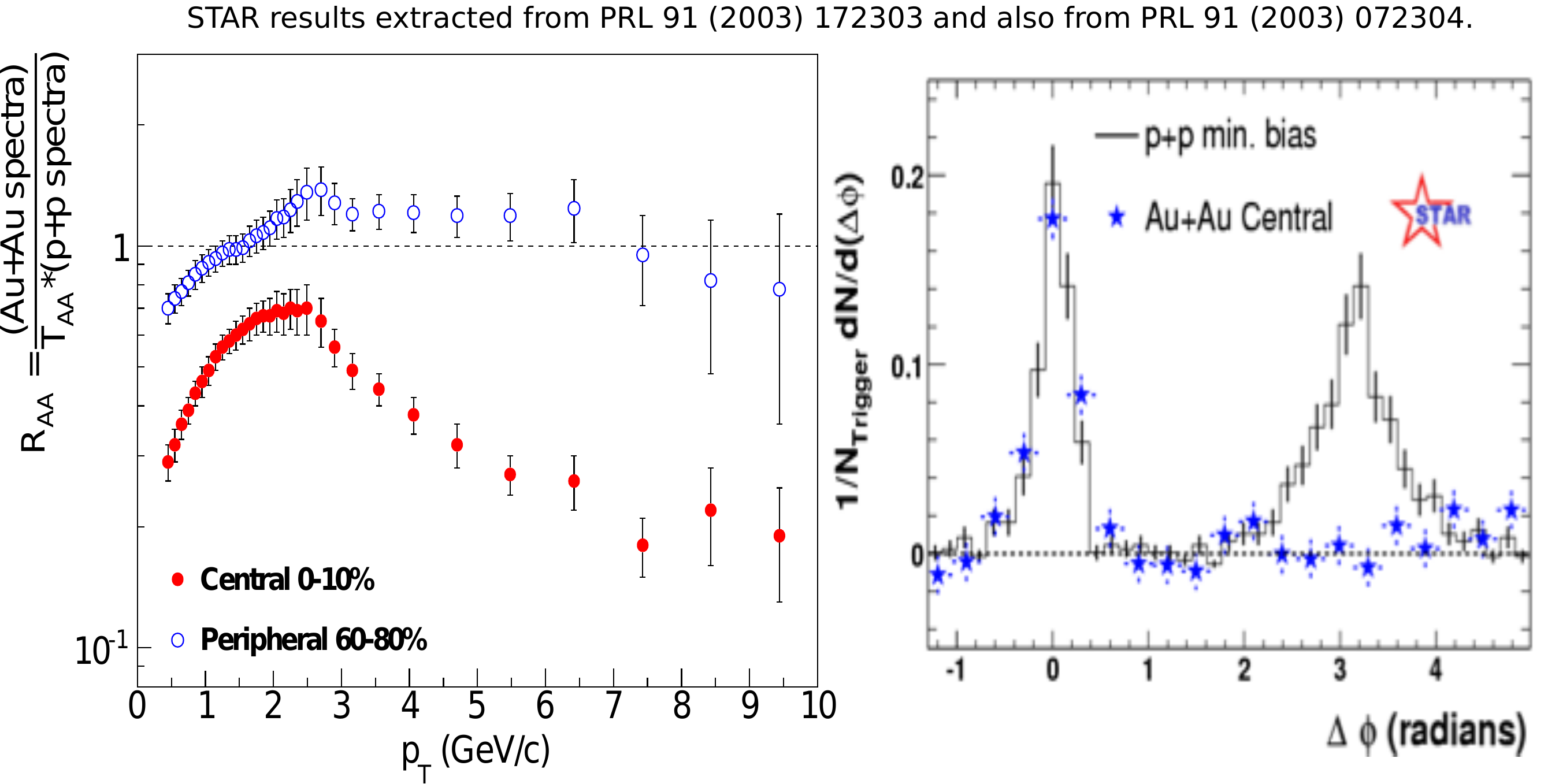}
\caption{Nuclear modification factor for peripheral (open blue circles) and central (solid red circles) for Au+Au collisions measured at 200 GeV. Plot was generated with data from~\cite{starRaa}. Right plot: Two particle azimuthal correlation function measured at central rapidity region of p+p collisions (black histogram) and central Au+Au collisions (blue star symbols), presented by the STAR experiment. Plot was generated with data from~\cite{starBBCorrelation}.}
\label{fig:5}
\end{figure}

\begin{figure}
\centering\includegraphics[width=.6\linewidth]{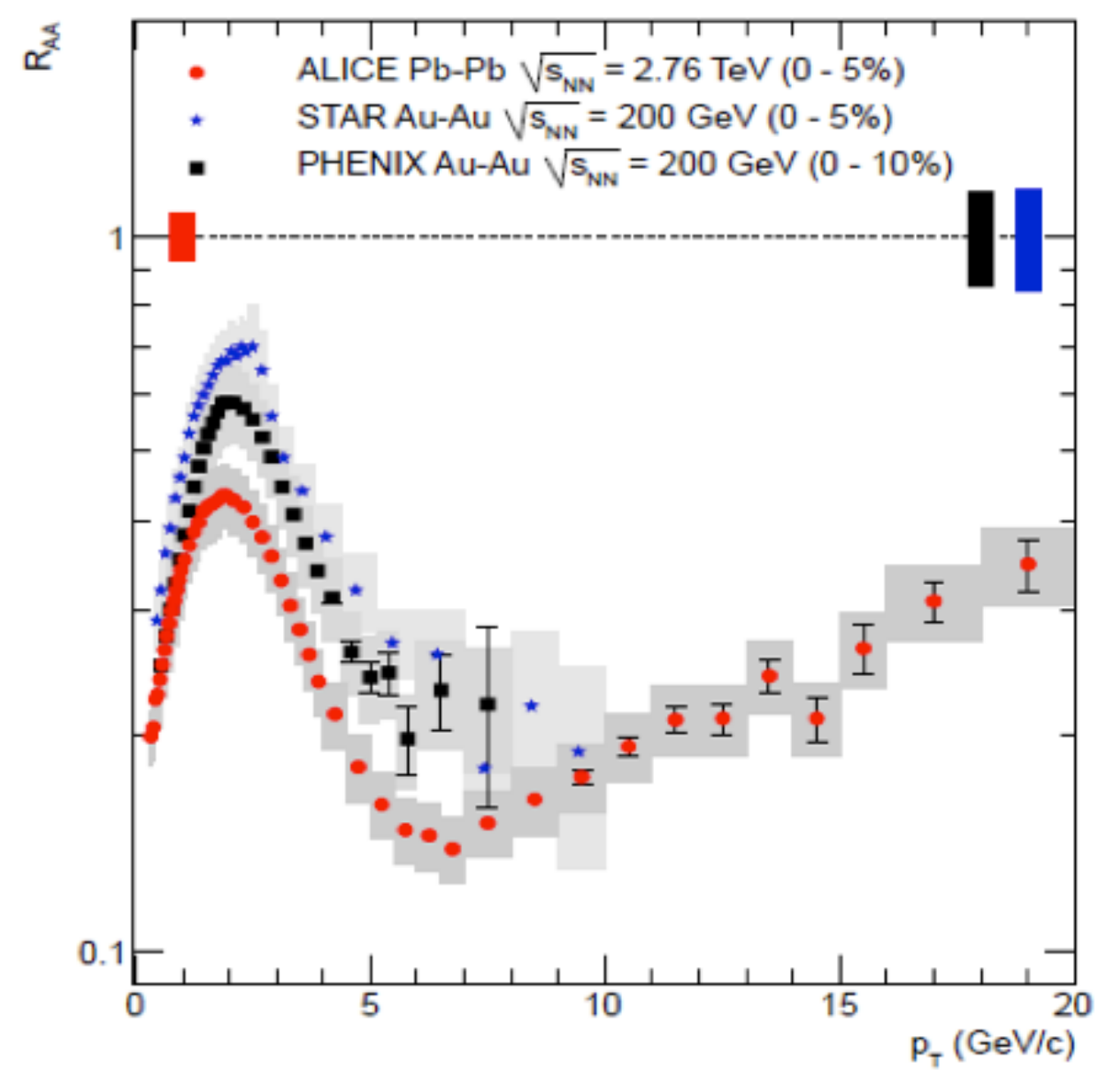}
\caption{Nuclear Modification factor plot from the ALICE experiment measured for Pb+Pb collisions at 2.76 TeV (red solid circles) compared to the STAR (blue stars) and PHENIX (black squares) results at 200 GeV. Plot was extracted from Ref.~\cite{aliceRaa}.}
\label{fig:6}
\end{figure}

Another surprising result from RHIC was the observation of signatures consistent with the formation of a system that exhibits collective behavior. The bulk of the produced particles are in the low $p_{T}$ region, where the spectrum seems to follow a typical exponential fall, similar to a Maxwell-Boltzmann distribution. The systematic behavior of the measured slope parameters from spectra of different particles is consistent with a rapidly expanding thermal source defined by a common kinetic freeze-out temperature and a common expansion velocity. Moreover, the relative abundance of the different types of particles is well described by statistical thermal models that assume chemical equilibration between the light quarks and even the strange quarks. There are different implementations of thermal models that consider total or partial equilibration of light and strange quarks and also grand-canonical or canonical ensembles~\cite{thermalModels,THERMUS}. But all models seem to describe reasonably well the different particle ratios, indicating some degree of equilibration. In figure 07, top panel, an example of the agreement between the measured ratios and the thermal model fit is presented, where data is from central Au+Au collisions measured by STAR and the horizontal bars are the fit results of the THERMUS code~\cite{THERMUS}. Figure 07 was extracted from Ref.~\cite{RDSPhD}. The chemical freeze-out temperature obtained from this fit of around 160 MeV is already very close to the limit predicted by Lattice QCD for the phase transition. In the lower panel of figure 07 we show a compilation of the chemical freeze-out temperatures and baryonic potential values obtained with the same thermal model applied to lower energy data. This kind of study is done to map the QCD phase diagram.

\begin{figure}
\centering\includegraphics[width=.6\linewidth]{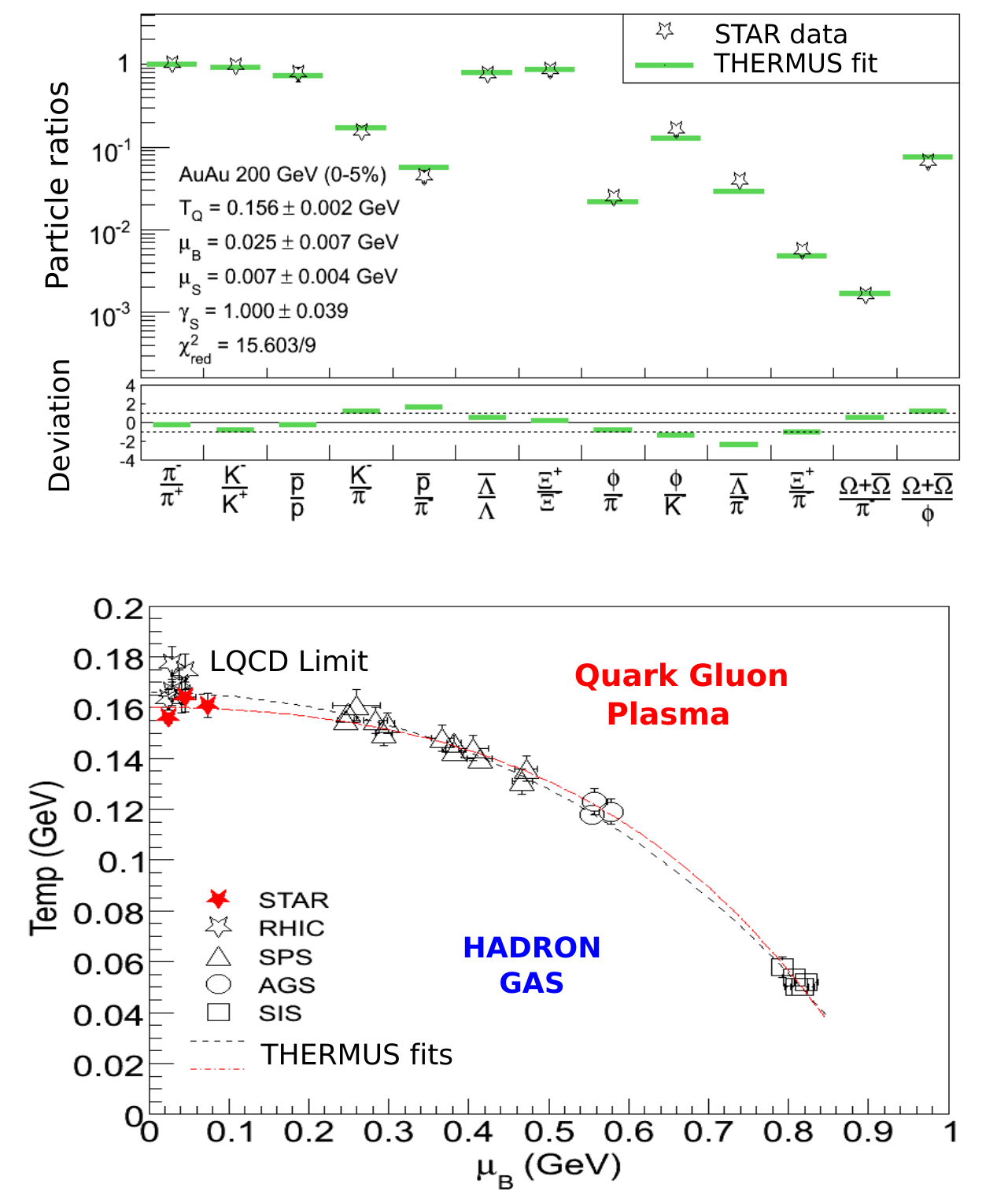}
\caption{Top panel: Example of the agreement between a statistical thermal model fit and experimentally measured particle ratios. Lower panel: Chemical Freeze-out temperature and Baryon Chemical potential from thermal model fits to various experimental results. Plots extracted from Ref.~\cite{RDSPhD} and the statistical thermal model used was the THERMUS code~\cite{THERMUS}.}
\label{fig:7}
\end{figure}

Perhaps, the most convincing signature of collectivity comes from the observable known as elliptic flow. In a non-central collision, the initial state of the nuclear collision has a non-symmetrical shape simply due to geometrical considerations. This geometrical anisotropy generates a difference in pressure that as the system evolves will develop into an anisotropy of the momentum distribution. By measuring the azimuthal distributions of particles (transverse to the beam direction), it is possible to de-convolute this distribution into a Fourier expansion. The amplitude of the second harmonic, v2, is known as the elliptic flow measurement. Within this picture, a non-zero measurement of the elliptic flow is consistent with a system that interacts strongly and shows collectivity through the evolution such that particles emitted in different directions are correlated. Large values of v2 could only be achieved if the system had very strong internal pressure and rapid thermalization so that there is enough time for elliptic flow to evolve. Hydrodynamical models describe well the general features of elliptic flow measurements, and models that consider for the evolution an equation of state with both partonic and hadronic stages seem to describe better the experimental values. Also, most hydodynamic models have no viscosity included in the evolution and still describe well the experimental results, suggesting thus, that the new state of matter formed exhibits very low viscosity values. In figure 08 left panel, we present the elliptic flow values measured by the STAR experiment for the different collision centrality classes, which was presented in Ref.~\cite{starPRL86}. In very peripheral collisions, due to the large initial geometrical asymmetry, a large value of elliptic flow is observed, and this anisotropy decreases as we go to more central collisions. Also, it is presented in the same figure the hydrodynamical limit for the elliptic flow measurement. More detailed comparisons, where new implementations of hydrodynamical models such as one that considers a fluctuating initial condition or new observables like momentum weighted correlation functions have been tested and compared to experimental data. New measurements from the LHC presented in Ref.~\cite{alicePRL105} by the ALICE experiment show a large value of v2 measured in Pb+Pb collisions at 2.76 TeV and also a decrease with centrality similar to the behavior observed at RHIC. The average integrated elliptic flow measured at 2.76 TeV is approximately $30\%$ higher than the values measured at RHIC, which is also higher than the predictions by hydrodynamical models with zero viscosity. Figure 08, right panels shows the evolution of the integrated v2 values for semi-peripheral collisions for the different collision energies. Plot was presented in Ref.~\cite{alicePRL105}. More detailed results combined with different models are still needed to understand the higher values of v2 measured at the LHC.

\begin{figure}
\centering\includegraphics[width=.9\linewidth]{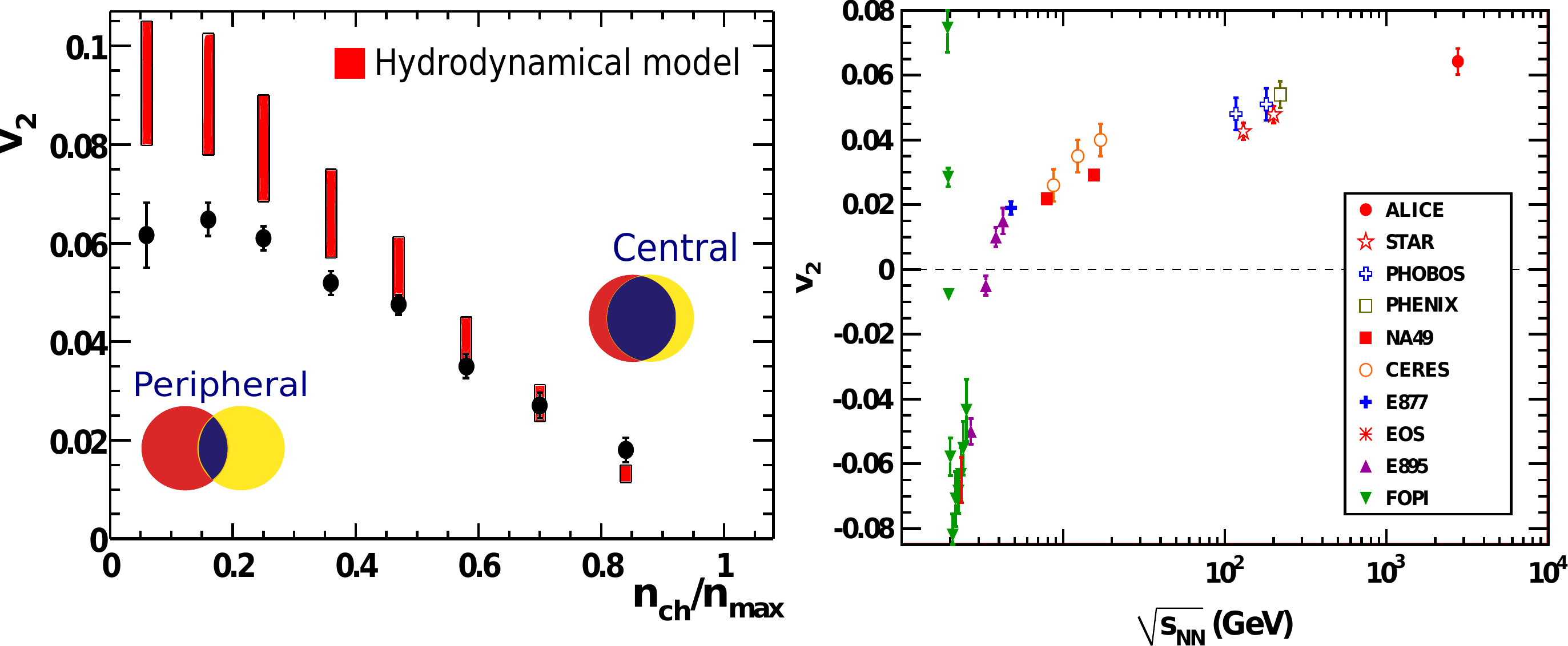}
\caption{Left panel: Elliptic flow values measured by the STAR experiment for the different collision centrality classes, which was presented in Ref.~\cite{starPRL86}. Right panel: Integrated elliptic flow values for semi-peripheral collisions as a function of collisions energy compiled from different experiments. Plot was presented in Ref.~\cite{alicePRL105} and shows the higher value measured by the ALICE experiment at 2.76 TeV.}
\label{fig:8}
\end{figure}

In addition to the search for signatures of the Quark Gluon Plasma and the characterization of the unique system formed in these high-energy nuclear collisions, the large amount of particles produced in these collisions also allowed the search for new and exotic particles. Recently, the discovery of an anti-hyper-triton, which is a bound nuclear state formed by an anti-proton, an anti-neutron and an anti-lambda, was announced and presented by the STAR collaboration in Ref.~\cite{starAHT}. Not much later, the same collaboration also announced the first observation of an anti-alpha nuclei, presented in Ref.~\cite{starAAlpha}, opening a new field of interest in relativistic heavy ion collisions. 
Data from two-particle correlation function in the non-central rapidity region showed an unexpected long-range correlation between particles in the longitudinal direction~\cite{starLRC}. Different explanations are being debated such as the existence of color flux-tubes in the longitudinal direction that could give rise to such correlations. These color flux tubes are proposed in a scenario where new states of matter such as Color Glass Condensate and Glasma would be created in relativistic heavy ion collisions. Within these proposed scenarios, the accelerated Lorentz boosted nuclei would form a sheet of high density gluons called Color Glass Condensate (CGC). As the collision happens, these sheets of gluons interact and form color flux tubes of color electric and color magnetic fields extending in the longitudinal direction. After some short time, these color flux tubes evolves into a pre-equilibrium phase called Glasma. The high color fields would then decay into gluons evolving into the strongly interacting Quark Gluon Plasma. A review of the proposed CGC and GLASMA scenario can be found at Ref.~\cite{CGC}. The experimentally measured long-range correlation was well described by a phenomenological model that considers these flux-tubes and the subsequent hydrodynamical evolution~\cite{myPRL,PhMota} but, the final confirmation of these flux tubes are still in debate. At the LHC, the CMS experiment has presented two particle correlation function from p+p collisions at 7 TeV that also shows a long-range correlation in the longitudinal direction ~\cite{CMSlongrange}. So, further studies and new ideas are still needed to complete and explain these measurements and test the proposed theories.

\section{The new frontier}

After more than a decade of RHIC operations, a large amount of data has been measured, with a wide variety of colliding nuclei (Au+Au, p+p, d+Au and Cu+Cu), with nucleon pair energies ranging from 10 GeV to 200 GeV. Due to the complexity of these collisions, where the system formed goes through different phases of matter and where several different mechanisms compete during the dynamical evolution, it is very difficult to single out a measurement that proves or falsify the existence of the Quark Gluon Plasma. However, the large amount of data and the many analysis tools developed over these last ten years allow for a systematic study of different physics observables and provide many signatures that are consistent with the formation of a state of matter with partonic degrees of freedom. From the results presented up to now we can safely conclude that central Au+Au or Pb+Pb collisions DO NOT behave like an incoherent superposition of p+p collisions and indeed a NEW STATE OF MATTER HAS BEEN FORMED.

At the LHC, with the higher collision energy, many probes that were of limited access at RHIC, such as heavy flavored particles and very high transverse momentum spectra, will become available. Up to now, the early results presented by the LHC experiments from the short run of Pb+Pb collisions at 2.76 TeV has already confirmed many of the features observed at RHIC, further enhancing the above statement that a new state of matter is indeed created in these collisions. But, more than just confirming, I expect that in a very short time, new measurements and new ideas will help us better understand the characteristics of this new state of matter and the features of the phase transition. Also, I hope that with this, we can finally start to extrapolate what we have learned in heavy-ion collisions to the cosmological models to understand what has happened in the very early stages of our universe, microseconds after the Big-Bang, and how it has evolved into the hadronic state of the Universe today. The study of new forms of matter at extreme conditions is still a very open field, both in the experimental and theoretical sides. It is very exciting to see many new results and new proposals of ideas and theories being presented every day.

\end{document}